\documentclass[aps,prb,amsmath,amssymb,amsfonts]{revtex4}
\usepackage[caption=false]{subfig}
\usepackage{graphicx}
\usepackage{color}
\usepackage{booktabs,array,multirow}

\begin{document}

\title{Can freestanding Xene monolayers behave as excitonic insulators?}
\author{Matthew N. Brunetti$^{1,2}$, Oleg L. Berman$^{1,2}$, and Roman Ya. Kezerashvili$^{1,2}$}
\affiliation{%
$^{1}$New York City College of Technology, City University of New York, Brooklyn, NY 11201, USA\\
$^{2}$The Graduate School and University Center,
City University of New York,
New York, NY 10016, USA
}

\date{\today}

\begin{abstract}
  We predict a phase transition in freestanding monolayer Xenes from the semiconducting phase to the excitonic insulating (EI) phase can be induced by reducing an external electric field below some critical value which is unique to each material.
  The splitting of the conduction and valence bands due to spin-orbit coupling at non-zero electric fields leads to the formation of $A$ and $B$ excitons in the larger or smaller band gap, with correspondingly larger or smaller binding energies.
  Our calculations show the coexistence of the semiconducting phase of $A$ excitons with the EI phase of $B$ excitons for a particular range of electric field.
  The dielectric environment precludes the existence of the EI phase in supported or encapsulated monolayer Xenes.
\end{abstract}

\maketitle

  In 1963, it was noted~\cite{Knox1963} that if the exciton binding energy exceeds the band gap in a semiconductor or semimetal, the material would behave as a so-called excitonic insulator (EI), where the valence electrons would spontaneously form excitons.
  Following Knox's remark, theoretical research on excitonic insulators grew in popularity over the course of the 1960s~\cite{Keldysh1965,DesC1965,Jerome1967,Keldysh1968}.
  Experimental evidence of EI phenomena has been found in semiconductor quantum wells such as InAs/GaSb~\cite{Du2017}, layered semiconductors such as Ta$_{2}$NiSe$_{5}$~\cite{Wakisaka2009}, and TMDCs such as 1\textit{T}-TiSe$_2$~\cite{Cercellier2007}.
  Most recently, theoretical research has suggested that previously anomalous experimental observations of carbon nanotubes may be evidence of the excitonic insulator phase~\cite{Rontani2014,Varsano2017}.
  These results motivated further research, especially in 2D materials following the success of the layered semiconductors Ta$_2$NiSe$_5$ and 1\textit{T}-TiSe$_2$.
  Around the same time, a number of theoretical studies predicted that freestanding (FS) graphene was a possible candidate for the EI phase~\cite{Drut2009,Reed2010,Stroucken2011,DeJuan2012,Stroucken2012}.
  Experimental studies of FS graphene performed around the same time showed convincing evidence that FS graphene in vacuum behaved as a semimetal, not an excitonic insulator~\cite{Eberlein2008,Gass2008}.
  Subsequent theoretical investigations using a refined theoretical model were in agreement with the aforementioned experimental results, namely that graphene should not exhibit an EI phase under any circumstances~\cite{Wang2017,Wang2012a,Popovici2013,Ulybyshev2013,Smith2014,Gonzalez2015,Carrington2016}.

  In recent years the buckled 2D allotropes of silicon (silicene), germanium (germanene), and tin (stanene) have seen a significant rise in research interest.
  It was theoretically predicted early on that buckled 2D materials (collectively, Xenes~\cite{Molle2017}) were unique amongst 2D materials because their band gaps could be tuned by an external electric field~\cite{Tabert2014,Ezawa2012,Ezawa2013a}.
  Other interesting phenomena in the Xenes include non-local plasmon modes in an open system~\cite{Iurov2017}, and the discovery of very strongly bound excitons in hydrogen-functionalized silicene and germanene~\cite{Pulci2012}.
  However, to this date it appears that the Xenes have not been considered as candidates for the excitonic insulator phase.
  There does not appear to be any experimental evidence of freestanding Xenes in the same way that experimentalists studied freestanding graphene from Refs.~\onlinecite{Eberlein2008} and~\onlinecite{Gass2008}, and hence it remains an open question as to whether or not the FS Xenes are excitonic insulators.

  A rigorous exploration of this question, one which might utilize the powerful theoretical approaches outlined by e.g. Ref.~\onlinecite{Wang2012a} is beyond the scope of this letter.
  We instead present straightforward calculations for the electric field-dependent exciton binding energy and band gap in the 2D Xenes.
  We show that, by using a common theoretical approach to model the dependence of the exciton binding energies and the band gaps on the external electric field, there appears to be enough evidence of the excitonic insulator phase in the FS Xenes to warrant further study.
  Any phase transition can be characterized by some internal critical parameter~\cite{Stanley1971}.
  Specifically, we demonstrate that when the external electric field is zero or less than some critical value which is unique to each material, the exciton binding energy is larger than the corresponding band gap.
  Therefore, according to Ref.~\onlinecite{Knox1963}, this is the condition which must be satisfied for the existence of the EI phase.
  In contrast, we also consider the Xenes on different substrates and find that the enhanced dielectric screening from the substrate drastically reduces the exciton binding energy such that the EI phase is no longer possible.


  In the Wannier-Mott model for excitons, the Schr\"{o}dinger equation for the electron and hole in an Xene monolayer reads:

  \begin{equation}
	\left[ - \frac{1}{2 m_e^*} \nabla_{\mathbf{r}_e}^2 - \frac{1}{2 m_h^*} \nabla_{\mathbf{r}_h}^2 + V(r) \right] \psi(\mathbf{r}) = E \psi(\mathbf{r}),
	\label{eq:ehschro}
  \end{equation}
  where $m_e^*$ and $m_h^*$ are the effective electron and hole masses, respectively, $\mathbf{r} = \mathbf{r}_e - \mathbf{r}_h$ is the relative position coordinate between the electron and hole in the Xene monolayer, and the potential $V(r)$ characterizes the electron-hole interation.
  The electrostatic interaction between the electron and hole is the Coulomb potential, $V_{C} = - (k e^2)/(\kappa r)$, where $k = 9 \times 10^9$ Nm$^2$/C$^2$, but in the case of a 2D layer in an inhomogeneous dielectric environment, the electron-hole interaction potential is modified by the dielectric screening of the Xene monolayer, as well as dielectric screening from its environment, such as the substrate upon which it is placed or within which it is encapsulated.
  Today, most calculations for 2D monolayers are performed by using the Rytova-Keldysh (RK) potential, which was first derived in Ref.~\onlinecite{Rytova1967} and re-derived 12 years later, in Ref.~\onlinecite{Keldysh1979}:

  \begin{equation}
	  V_{RK} (r) = - \left( \frac{\pi k e^2}{2 \kappa \rho_0} \right) \left[ H_0 \left( \frac{r}{\rho_0} \right) - Y_0 \left( \frac{r}{\rho_0} \right) \right].
	\label{eq:rkpot}
  \end{equation}
  In Eq.~\eqref{eq:rkpot}, $\kappa = (\epsilon_1 + \epsilon_2)/2$, where $\epsilon_1$ and $\epsilon_2$ correspond to the dielectric constant of the environment above and below the Xene monolayer, respectively, $\rho_0 = l \epsilon / (\epsilon_1 + \epsilon_2)$ is the so-called dielectric screening length, where $l$ is the Xene monolayer thickness and $\epsilon$ is the bulk dielectric constant of the Xene, and $H_0$ and $Y_0$ are the Struve and Bessel functions of the second kind, respectively.
  The asymptotic behavior of the RK potential is given by:

  \begin{equation}
	V_{RK} (r) = \begin{cases}
	  \frac{k e^2}{\kappa \rho_0} \left[ \ln \left( \frac{r}{2 \rho_0} \right) + \gamma \right]	&	r \ll \rho_0 \vspace{0.1cm} \\
	  - \frac{k e^2}{\kappa r}	&	r \gg \rho_0
	\end{cases}.
	\label{eq:rkasymptotic}
  \end{equation}
  where $\gamma$ is Euler's constant.
  At the same time this potential for the small screening length $\rho_0$ reduces to the bare Coulomb potential, while for very large screening lengths $\rho_0$ the interaction potential becomes logarithmic.
  In our calculations we consider the Xene monolayer as suspended in vacuum, supported on a dielectric substrate, and encapsulated by a dielectric, using both the Coulomb and Rytova-Keldysh potentials, so as to comprehensively account for the effect of dielectric screening on the system.

  It was found that the band gap in Xene monolayers can be affected by an external electric field perpendicular to the Xene layer.
  The theoretical approach describing the band gaps, $2 \Delta_{\xi\sigma}$, carrier masses, $m^*$, and their dependence on an external electric field $E_\perp$ in an Xene monolayer are presented in Refs.~\onlinecite{Drummond2012,Ezawa2012,Ezawa2013a,Tabert2014,Fadaie2016}.
  The electric field-dependent band gap in the Xenes is given by:
  \begin{equation}
	\Delta_{\xi\sigma} = \lvert \xi\sigma \Delta_{gap} - e d_0 E_\perp \rvert,
	\label{eq:gapez}
  \end{equation}
  where $\xi = \sigma = \pm 1$ are the valley and spin indices, respectively, $\Delta_{gap}$ is the intrinsic band gap of the Xene, $d_0$ is the buckling parameter, which gives the vertical offset between the two triangular Xene sublattices, and $E_\perp$ is the external electric field.
  Note that $\Delta_{\xi\sigma}$ represents the energy gap between the Fermi energy and either the bottom of the conduction band or the top of the valence band.
  Therefore, the full band gap between the valence and conduction bands is given by $2 \Delta_{\xi\sigma}$.
  Furthermore, there are two conduction bands and two valence bands denoted by combinations of $\xi$ and $\sigma$ which yield either a larger band gap, $\xi\sigma=-1$, or a smaller band gap, $\xi\sigma=1$, and therefore a larger or smaller effective carrier mass, respectively.
  Excitons formed in the larger band gap are referred to as $A$ excitons, while excitons from the smaller gap are $B$ excitons.
  The conduction and valence bands denoted by $\xi\sigma=-1$ will always move apart from each other under an external electric field.
  The conduction and valence bands corresponding to $\xi\sigma=1$ move towards each other as the electric field is increased from zero, form a Dirac cone for some particular value of the electric field, which we call the zero-gap field, $E_{\perp}^{0}$, and as the electric field increases beyond $E_{\perp}^{0}$, the bands will separate from each other.
  The values of the zero-gap electric fields are $E_{\perp}^{0} =$ 2.06 mV/\AA, 24.41 mV/\AA, 59.41 mV/\AA~in FS Si, Ge, and Sn, respectively.
  At $E_{\perp}^{0}$, the charge carriers are described by the Weyl equation, and because the band gap is zero at this value of electric field, no $B$ excitons can form.
  However, in our calculations we present results for the binding energies obtained for the RK potential, $E_{b}$, at values of $E_\perp$ which are very close to $E_{\perp}^{0}$.

  The dispersion relation of the bands is given by

  \begin{equation}
	E(\mathbf{p}) = \sqrt{\Delta_{\xi\sigma}^{2} + v_F^2 \lvert \mathbf{p} \rvert^{2}},
	\label{eq:disprelstandard}
  \end{equation}
  where $v_F$ is the Fermi velocity of charge carriers.
  Assuming $\lvert \mathbf{p} \rvert$ is small compared to the total band gap $\Delta_{\xi\sigma}$, one can expand Eq.~\eqref{eq:disprelstandard} in terms of $\lvert \mathbf{p} \rvert^2$, and obtain

  \begin{equation}
	E(\mathbf{p}) \approx \Delta_{\xi\sigma} + \frac{v_F^2 \lvert \mathbf{p} \rvert^2}{2 \Delta_{\xi\sigma}}.
	\label{eq:disprelexpand}
  \end{equation}
  In Eq.~\eqref{eq:disprelexpand}, the second term can be interpreted as the kinetic energy of charge carriers with effective mass $m^* = \Delta_{\xi\sigma}/v_F^2$.
  Therefore, recalling Eq.~\eqref{eq:gapez}, one obtains~\cite{Drummond2012,Ezawa2012,Ezawa2013a,Tabert2014,Fadaie2016}:

  \begin{equation}
	m^* = \frac{\lvert \xi\sigma\Delta_{gap} - e d_0 E_{\perp} \rvert}{v_F^2}.
	\label{eq:effmassez}
  \end{equation}

  In the case when the electron and hole are bound by the Coulomb potential, one can obtain an analytical expression for the Wannier-Mott exciton binding energy as a function of external electric field:
  \begin{equation}
	E_b^{C}(E_\perp) = \frac{2 k^2 e^4}{v_F^2 \hbar^2 \kappa^2} \times \lvert \xi\sigma \Delta_{gap} - e d_0 E_\perp \rvert.
	\label{eq:ebcoulez}
  \end{equation}

  Therefore, the ratio $E_{b}^{C}/(2\Delta_{\xi\sigma})$ is a constant, which is different for each material \textendash{} specifically, $E_{b}^{C}/(2 \Delta_{\xi\sigma}) = 5.66,~6.23,~7.91$ for freestanding Si, Ge, and Sn, respectively.
  However, if one imagines an Xene monolayer encapsulated by hexagonal boron nitride ($h$-BN) with $\kappa = \epsilon_{h-\text{BN}} = 4.89$, the value of the ratio of binding energy to band gap becomes $0.237,~0.260,~0.330$, respectively, while when the Xenes are placed on an SiO$_{2}$ substrate in vacuum, the corresponding ratios are 0.982, 1.08, and 1.37, respectively.
  These results demonstrate that the excitonic insulator phase is very sensitive to the dielectric environment.


  \begin{table*}[t]
	\centering
	\begin{tabular}{|r|c|c|c|c|c|}
	  \hline
	  Material		&	$2\Delta_{\text{gap}}$ (meV)&	$d_0$ (\AA)					&	$v_F$ ($\times 10^5~\text{m/s}$)	&	$\epsilon$	&	$l$ [nm]\\\hline
	  Silicene		&	1.9~\cite{Matthes2013}		&	0.46~\cite{Ni2012}			&	6.5~\cite{Matthes2013}				&	11.9	&	0.4~\cite{Tao2015}\\\hline
	  Germanene		&	33~\cite{Matthes2013}		&	0.676~\cite{Ni2012}			&	6.2~\cite{Matthes2013}				&	16		&	0.45\\\hline
	  Stanene		&	101~\cite{Matthes2013}		&	0.85~\cite{Balendhran2015}	&	5.5~\cite{Matthes2013}				&	24		&	0.5\\\hline
	\end{tabular}
	\caption{\label{tab:matpars}%
	  Parameters for buckled 2D materials:
	  $2\Delta_{\text{gap}}$ is the total gap between the conduction and valence bands at $E_\perp = 0$,
	  $d_0$ is the buckling parameter,
	  $v_F$ is the Fermi velocity,
	  $l$ is the monolayer thickness, and
	  $\epsilon$ is the dielectric constant of the bulk material.
	}
  \end{table*}

\begin{figure}[h]
  \centering
  \includegraphics[width=0.49\columnwidth]{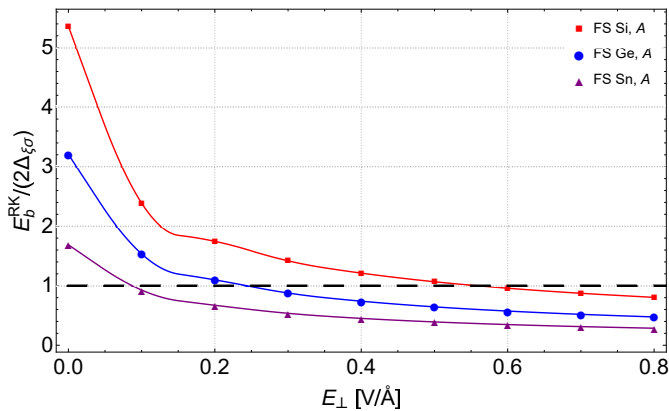}
  \caption{
	Dependence of the ratio of $A$ exciton binding energy to bandgap, $E_b/(2\Delta_{\xi\sigma})$, on the external electric field, $E_\perp$, in the freestanding Xenes.
	The dashed line corresponds to $E_b / (2 \Delta_{\xi\sigma}) = 1$.
  }
  \label{fig:ebratio}
\end{figure}

  Let us now consider the formation of the Wannier-Mott excitons via the RK potential.
  We calculated the binding energies of excitons in the freestanding Xenes when $E_{\perp} = 0$ V/\AA.
  The values for the binding energies of excitons at zero electric field are $E_{b} = 10.1~\text{meV},~105.9~\text{meV},~\text{and}~170.4$ meV in Si, Ge, and Sn, respectively.
  Comparing these values to the intrinsic band gaps in the three Xenes found in Table~\ref{tab:matpars} and given by $2 \Delta_{\text{gap}} = 1.9~\text{meV},~33~\text{meV},~\text{and}~101$ meV, respectively, one observes that these binding energies are far larger than their respective band gaps, which could be an indicator of the EI phase in these materials.

  Fig.~\ref{fig:ebratio} shows the dependence of the ratio of the exciton binding energy, calculated using the RK potential for $A$ excitons, to the band gap, $E_b/(2 \Delta_{\xi\sigma})$, on the external electric field, $E_\perp$.
  First, we find that FS Si would behave as an EI until the external electric field exceeds the critical value of about $E_{\perp c} = 0.55~\text{V/\AA}$, where the critical electric field $E_{\perp c}$ is defined as the electric field at which $E_b / (2 \Delta_{\xi\sigma}) = 1$.
  The corresponding critical fields in FS Ge and FS Sn are approximately $E_{\perp c} = 0.3~\text{V/\AA}$ and $E_{\perp c} = 0.2~\text{V/\AA}$, respectively.

\begin{figure}[h]
  \centering
  \includegraphics[width=0.49\columnwidth]{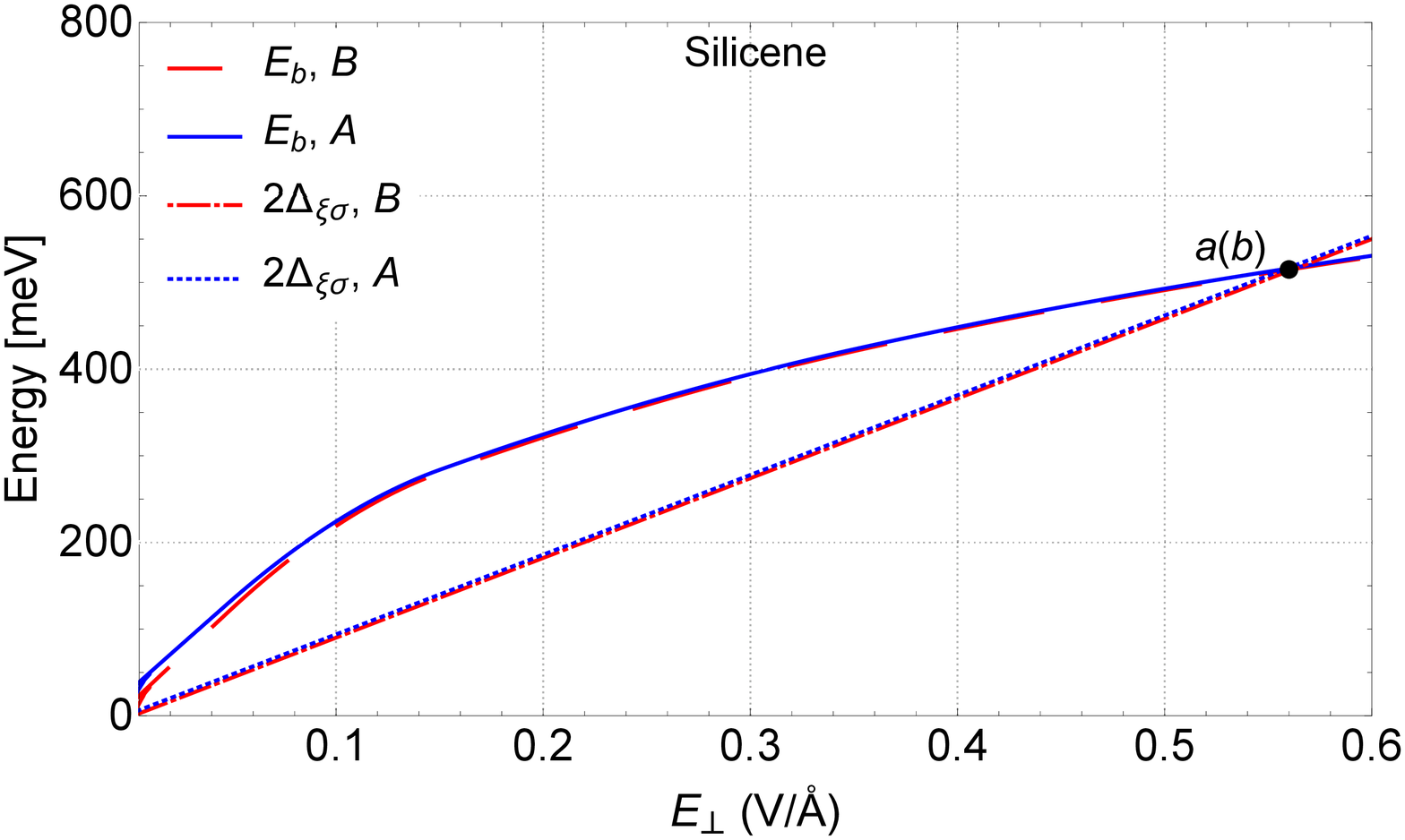}
  \includegraphics[width=0.49\columnwidth]{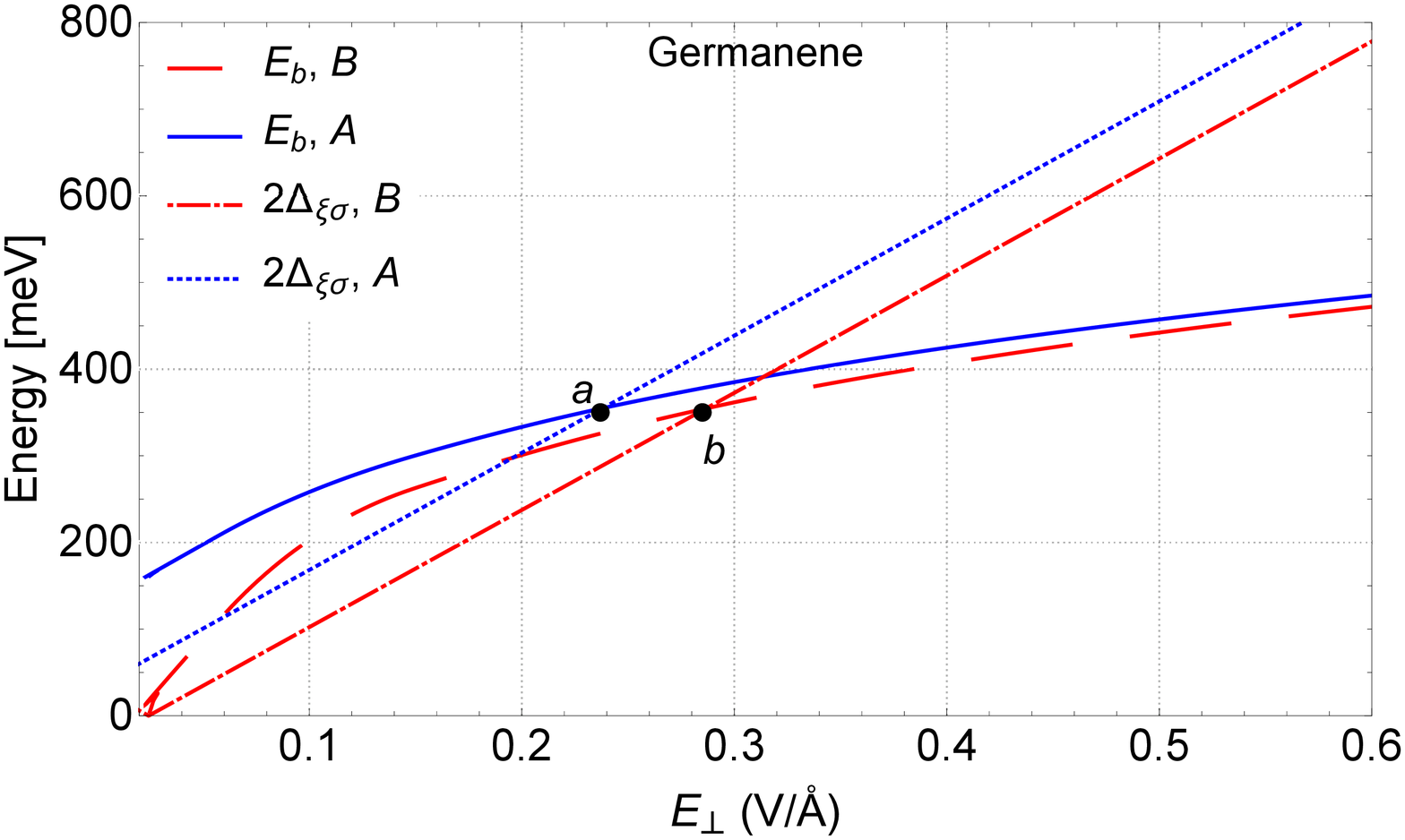}
  \includegraphics[width=0.49\columnwidth]{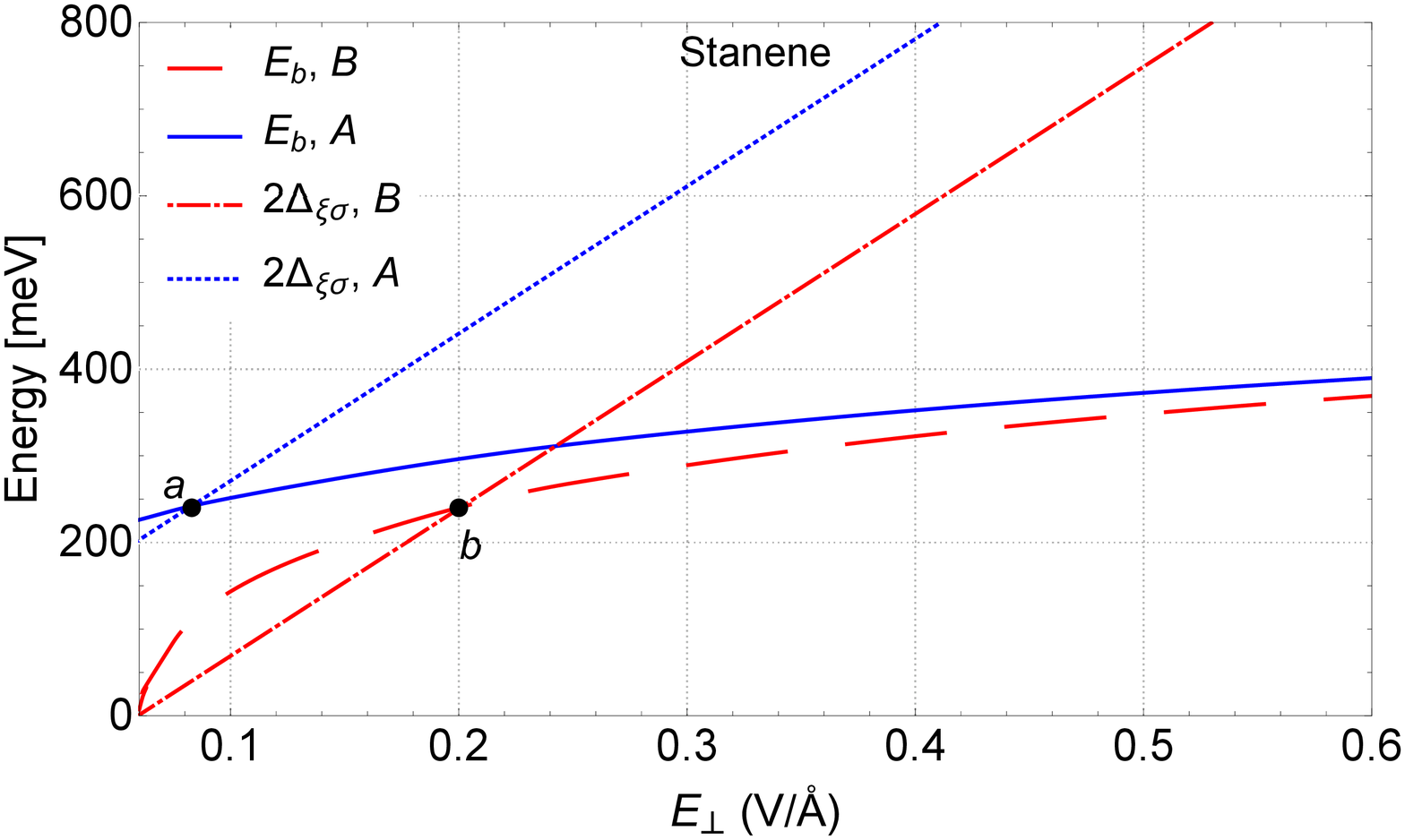}
  \caption{
	The dependence of exciton binding energy, $E_b$, and band gap, $2 \Delta_{\xi\sigma}$, for $A$ excitons (blue) and $B$ excitons (red) on the external electric field, $E_\perp$, in freestanding silicene, germanene, and stanene.
	The black dots show the point at which the band gap and corresponding binding energy are equal to each other for the two types of excitons.
	A phase transition is expected to occur at the value of the electric field corresponding to the points labeled $a$ for $A$ excitons and $b$ for $B$ excitons.
	Calculations performed for the RK potential.
  }
  \label{fig:ebgap}
\end{figure}

  Fig.~\ref{fig:ebgap} shows both the binding energy obtained by using the RK potential, $E_b$, and the band gap, $2 \Delta_{\xi\sigma}$, as a function of external electric field.
  The points where the binding energy curve crosses over the band gap line, denoted by $a$ for $A$ excitons and $b$ for $B$ excitons, correspond to the phase transitions between the EI and semiconducting phases.
  Since the band gap corresponding to $A$ excitons is by definition larger than the gap corresponding to $B$ excitons, the phase transition for $A$ excitons always occurs at a lower value of $E_{\perp}$ than the phase transition for $B$ excitons.
  The tunability of the EI phase is illustrated well here.
  Since the dependence of the band gap on the electric field is linear, while the dependence of $E_b$ on the band gap is monotonically increasing but non-linear (upon inspection, of order less than 1), the difference between the exciton binding energy and the band gap varies drastically and in a non-linear way, allowing for a great deal of freedom and creativity if one would need to tune the EI band gap to a particular value.
  Our calculations also show the coexistence of the semiconductor and excitonic insulator phases.
  Especially in Sn, but also in Ge, we are able to see that there is a range of electric field for which the $A$ excitons are in the semiconducting phase, while the $B$ excitons are in the EI phase.
  Thus, in this range of electric field, the semiconducting phase of $A$ excitons coexists with the EI phase of $B$ excitons.
  The experimental observation of these two phases coexisting in a single monolayer could have profound implications on the future of nanodevice design and utilization.

  In the case of the Coulomb potential, the excitonic insulator phase is formed in freestanding Xenes for any electric field, provided one chooses $\kappa = 1$, corresponding to the freestanding Xene monolayers in vacuum.
  The Coulomb binding energy scales inversely with the square of the environmental dielectric constant $\kappa$, so that on certain substrates the excitonic insulator phase is not formed and excitons can be created by laser pumping as optical excitations above the semiconducting ground state, for any eletric field.

  \begin{figure}[h]
	\centering
	\includegraphics[width=0.5\columnwidth]{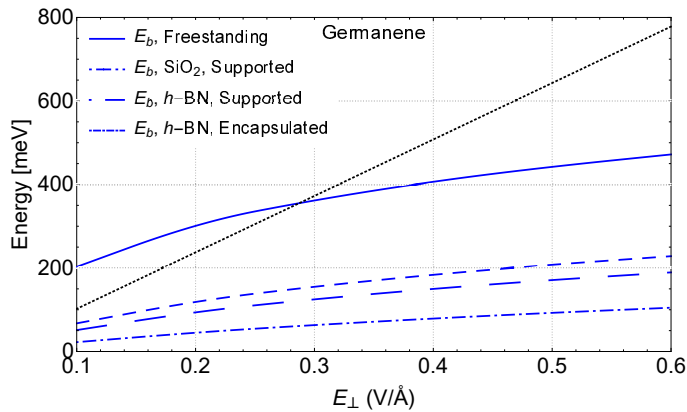}
	\caption{
	  Binding energy of $B$ excitons in Ge on various substrates, along with the electric field dependent band gap, as a function of external electric field.
	  Calculations performed for the RK potential.
	  The band gap in germanene, which does not depend on the dielectric environment, is shown by the black dashed line.
	}
	\label{fig:gesubstrates}
  \end{figure}

  To understand the importance of dielectric screening on the formation of the EI phase, we perform calculations of the exciton binding energy for Xenes placed in three different dielectric environments: supported on an SiO$_2$ $(\epsilon_{\text{SiO}_2} = 3.8)$~\cite{VanDerDonck2017} substrate, supported on an $h$-BN $(\epsilon_{h-\text{BN}}=4.89)$~\cite{Fogler2014} substrate, and encapsulated by $h$-BN, and compare these results to the freestanding binding energy as well as the electric field-dependent band gap, which is independent of the dielectric environment.
  It should be mentioned that the parameters presented in Table~\ref{tab:matpars} are obtained for freestanding monolayer Xenes.
  In Refs.~\onlinecite{Li2013} and~\onlinecite{Kaloni2013a} the authors studied silicene on an $h$-BN substrate using \textit{ab initio} calculations and arrived at completely different results related to the changes of the material parameters of a silicene monolayer due to interaction with the substrate.
  In Ref.~\onlinecite{Brunetti2018b}, using the parameters for silicene on an $h$-BN substrate we calculated the binding energies for silicene encapsulated by $h$-BN.
  Analysis of these results led to the conclusion that the EI phase is not possible in silicene encapsulated by $h$-BN, as evidenced by our calculation which found that the ratio $E_b / (2 \Delta_{\xi\sigma}) < 1$ for all values of $E_\perp$.
  On the other hand, the parameters for germanene and stanene on an $h$-BN substrate or encapsulated by $h$-BN are not available.
  Therefore, we present the aforementioned calculations of the Xenes on different substrates with the acknowledgement that the material parameters of Ge and Sn may change drastically when placed on a substrate.

  Fig.~\ref{fig:gesubstrates} shows the result of these calculations for $B$ excitons in germanene for various dielectric environments.
  It can be easily seen that the dielectric environment has a significant effect on the exciton binding energy, such that the calculations show that the EI phase is not possible in germanene when placed on any substrate.
  The ratio of binding energy to band gap, $E_b / \left( 2 \Delta_{\xi\sigma} \right)$ is largest for the freestanding Xenes and gets progressively smaller as the dielectric screening from the environment increases, where the ratio is smaller for an Xene supported on $h$-BN than on SiO$_2$, and is the smallest for $h$-BN encapsulation.
  It is worth mentioning that the corresponding calculations for stanene show that $B$ excitons in stanene supported by SiO$_2$ may still exhibit EI behavior at electric fields which are very close to $E_{\perp}^0$.
  It is important to note that the exciton binding energy and band gaps are very small in this regime, and thus the EI phase, while technically possible in this scenario, may not be practical.

  For the RK potential the ratio of binding energy to band gap depends on the electric field, and due to the transformation of the ground state the excitonic insulator is formed only at certain electric fields for freestanding Xenes, while for larger electric fields the excitons are created by laser pumping as optical excitations.
  In general, the excitonic insulator phase is formed due to strong electron-hole interactions and weak dielectric screening.
  The screening effects strongly suppress the electron-hole attraction and, therefore, prevent the spontaneous formation of the excitonic insulator state.
  In the Coulomb potential only the dielectric screening by the environment is considered while the RK potential more accurately accounts for both the dielectric constant of the environment as well as the dielectric screening provided by the 2D Xene monolayer.

  Our calculations demonstrate that for the Coulomb potential in freestanding Xenes the ratio of the exciton binding energy to the gap does not depend on electric field and is always greater than one.
  This is because the Coulomb binding energy depends linearly on the exciton reduced mass, which in turn depends linearly on the electric field, just as the band gap does.
  Therefore, for the Coulomb potential the excitonic insulator phase is formed in freestanding Xenes for any electric field.
  However, for the Coulomb potential in encapsulated Xenes embedded in certain dielectrics this ratio is less than one and does not depend on electric field.
  Therefore, for the Coulomb potential in encapsulated Xenes embedded in the certain dielectrics the excitonic insulator is not formed and excitons can be created by laser pumping as optical excitations above the ground state an any electric field.

  For the RK potential the ratio of exciton binding energy to the band gap depends on electric field, since the exciton binding energy calculated using the RK potential increases monotonically, but non-linearly, as the electric field (and therefore, the exciton reduced mass) is increased, and therefore the excitonic insulator phase is only observed at small electric fields for freestanding Xenes, while for larger electric fields, when the Xene monolayer is in the semiconducting phase, excitons are created by laser pumping as optical excitation.
  In general, the excitonic insulator phase is formed due to strong electron-hole interactions.
  Enhanced dielectric screening suppresses the electron-hole interaction and, therefore, prevents the formation of the excitonic insulator phase.
  The screening effects are taken into account either by the dielectric constant in Coulomb potential for the encapsulated Xenes or by employing the RK potential for both freestanding and encapsulated Xenes.

  At larger electric fields the ground state of freestanding Xenes will be a semiconductor, which conducts an electric current at finite temperatures.
  Without optical excitation by laser pumping, excitons are absent in Xenes with semiconducting ground state.
  When the electric field is reduced, there is a phase transition associated with transformation of the ground state resulting in the formation of an excitonic insulator.
  In the excitonic insulator phase the excitons are formed spontaneously (not due to optical excitation by laser pumping).
  The excitonic insulator phase is formed due to spontaneously broken symmetry.
  Since the excitonic insulator is formed by electrically neutral bound particles (excitons), the excitonic insulator does not conduct an electric current.
  The phase transition from the semiconductor to the excitonic insulator phase can be observed by a drop in electric current.


  We predict that a phase transition in freestanding Xene monolayers from the semiconducting phase to the insulating phase can be induced by reducing an external electric field below some critical value which is unique to each material.
  Furthermore, our calculations show that the excitonic insulator phase in freestanding Xenes is the ground state when the external electric field is zero, and that the semiconducting phase is accessed by increasing the external electric field beyond the critical value.
  We also find that the existence of the excitonic insulator phase is extremely sensitive to the dielectric environment, such that the excitonic insulator phase is destroyed by dielectric screening due to, for example, encapsulation of the Xene monolayer by $h$-BN.

	Based on the predicted phase transition at the material-specific critical electric field $E_{\perp c}$, we propose that an electric current switch could be designed using any of these materials, where electric current is turned off for $E_\perp < E_{\perp c}$ (excitonic insulator phase) and turned on for $E_{\perp} > E_{\perp c}$ (semiconductor phase).
	At large electric fields, freestanding Xenes behave as conductors in the semiconducting ground state when the switch is on, while for small electric fields Xenes behave as insulators in the excitonic insulator ground state, when the switch is off.

  Let us remark that the unique tunability of the band gap in buckled 2D crystals via an electric field allows us to explore the possibility of the EI phase in these materials.
  Based on our current understanding, this phenomenon would not be conceivable in other 2D materials such as, for example, the transition metal dichalcogenides (TMDCs) or phosphorene.
  Both the TMDCs~\cite{Kormanyos2015} and phosphorene~\cite{Tran2014,Liu2014a} exhibit large exciton binding energies due to their reduced dimensionality, though a recent theoretical paper~\cite{Trushin2016} showed that exciton binding energies in TMDCs are smaller than previously predicted.
  The main reason why TMDCs and phosphorene cannot be excitonic insulators is due to their very large band gaps \textendash{} in excess of 2 eV, while the exciton binding energy is a small fraction of the band gap.
  However, if one could find a mechanism which either reduces the band gap or increases the exciton binding energy such that $E_{bind} - E_{gap} > 0$ is satisfied, then one could expect to observe the EI phase in these materials.

  Our calculations show that a phase transition from the semiconductor phase to the excitonic insulator phase should occur in the freestanding Xenes as the electric field is reduced below some critical value, $E_{\perp c}$.
  In addition, we find that the freestanding Xenes should remain in the excitonic insulator phase even in the absence of an external electric field.
  Interestingly, our calculations show that the critical electric field $E_{\perp c}$ is different for $A$ and $B$ excitons, leading to a range of electric field for which the $A$ exciton in the semiconducting phase coexists with the $B$ exciton in the excitonic insulator phase.
  Further calculations which consider the Xenes on various substrates shows that the EI phase is extremely sensitive to the dielectric environment \textendash{} with the possible exception of $B$ excitons in stanene on an SiO$_2$ substrate, we find that the presence of any substrate is enough to decrease the exciton binding energy such that it is always smaller than the band gap, precluding the existence of the EI phase in these scenarios.

  \section*{Acknowledgements}

  The authors are grateful to Yurii E. Lozovik for the insightful conversation.
  This work is supported by U.S. Department of Defense under Grant No. W911NF1810433.


\begin{thebibliography}{48}
\expandafter\ifx\csname natexlab\endcsname\relax\def\natexlab#1{#1}\fi
\expandafter\ifx\csname bibnamefont\endcsname\relax
  \def\bibnamefont#1{#1}\fi
\expandafter\ifx\csname bibfnamefont\endcsname\relax
  \def\bibfnamefont#1{#1}\fi
\expandafter\ifx\csname citenamefont\endcsname\relax
  \def\citenamefont#1{#1}\fi
\expandafter\ifx\csname url\endcsname\relax
  \def\url#1{\texttt{#1}}\fi
\expandafter\ifx\csname urlprefix\endcsname\relax\def\urlprefix{URL }\fi
\providecommand{\bibinfo}[2]{#2}
\providecommand{\eprint}[2][]{\url{#2}}

\bibitem[{\citenamefont{Knox}(1963)}]{Knox1963}
\bibinfo{author}{\bibfnamefont{R.~S.} \bibnamefont{Knox}}, in
  \emph{\bibinfo{booktitle}{Solid State Physics}}, edited by
  \bibinfo{editor}{\bibfnamefont{F.}~\bibnamefont{Seitz}} \bibnamefont{and}
  \bibinfo{editor}{\bibfnamefont{D.}~\bibnamefont{Turnbull}}
  (\bibinfo{publisher}{Academic Press Inc.}, \bibinfo{address}{New York},
  \bibinfo{year}{1963}).

\bibitem[{\citenamefont{Keldysh and Kopaev}(1965)}]{Keldysh1965}
\bibinfo{author}{\bibfnamefont{L.~V.} \bibnamefont{Keldysh}} \bibnamefont{and}
  \bibinfo{author}{\bibfnamefont{{\relax{Yu}}.~V.} \bibnamefont{Kopaev}},
  \bibinfo{journal}{Sov. Phys. Solid State} \textbf{\bibinfo{volume}{6}},
  \bibinfo{pages}{2219} (\bibinfo{year}{1965}).

\bibitem[{\citenamefont{Des~Cloizeaux}(1965)}]{DesC1965}
\bibinfo{author}{\bibfnamefont{J.}~\bibnamefont{Des~Cloizeaux}},
  \bibinfo{journal}{J. Phys. Chem. Solids} \textbf{\bibinfo{volume}{26}},
  \bibinfo{pages}{259} (\bibinfo{year}{1965}).

\bibitem[{\citenamefont{J{\'{e}}rome et~al.}(1967)\citenamefont{J{\'{e}}rome,
  Rice, and Kohn}}]{Jerome1967}
\bibinfo{author}{\bibfnamefont{D.}~\bibnamefont{J{\'{e}}rome}},
  \bibinfo{author}{\bibfnamefont{T.~M.} \bibnamefont{Rice}}, \bibnamefont{and}
  \bibinfo{author}{\bibfnamefont{W.}~\bibnamefont{Kohn}},
  \bibinfo{journal}{Phys. Rev.} \textbf{\bibinfo{volume}{158}},
  \bibinfo{pages}{462} (\bibinfo{year}{1967}).

\bibitem[{\citenamefont{Keldysh and Kozlov}(1968)}]{Keldysh1968}
\bibinfo{author}{\bibfnamefont{L.~V.} \bibnamefont{Keldysh}} \bibnamefont{and}
  \bibinfo{author}{\bibfnamefont{A.~N.} \bibnamefont{Kozlov}},
  \bibinfo{journal}{Sov. Phys. JETP} \textbf{\bibinfo{volume}{27}},
  \bibinfo{pages}{521} (\bibinfo{year}{1968}).

\bibitem[{\citenamefont{Du et~al.}(2017)\citenamefont{Du, Li, Lou, Sullivan,
  Chang, Kono, and Du}}]{Du2017}
\bibinfo{author}{\bibfnamefont{L.}~\bibnamefont{Du}},
  \bibinfo{author}{\bibfnamefont{X.}~\bibnamefont{Li}},
  \bibinfo{author}{\bibfnamefont{W.}~\bibnamefont{Lou}},
  \bibinfo{author}{\bibfnamefont{G.}~\bibnamefont{Sullivan}},
  \bibinfo{author}{\bibfnamefont{K.}~\bibnamefont{Chang}},
  \bibinfo{author}{\bibfnamefont{J.}~\bibnamefont{Kono}}, \bibnamefont{and}
  \bibinfo{author}{\bibfnamefont{R.~R.} \bibnamefont{Du}},
  \bibinfo{journal}{Nat. Commun.} \textbf{\bibinfo{volume}{8}},
  \bibinfo{pages}{1971} (\bibinfo{year}{2017}).

\bibitem[{\citenamefont{Wakisaka et~al.}(2009)\citenamefont{Wakisaka, Sudayama,
  Takubo, Mizokawa, Arita, Namatame, Taniguchi, Katayama, Nohara, and
  Takagi}}]{Wakisaka2009}
\bibinfo{author}{\bibfnamefont{Y.}~\bibnamefont{Wakisaka}},
  \bibinfo{author}{\bibfnamefont{T.}~\bibnamefont{Sudayama}},
  \bibinfo{author}{\bibfnamefont{K.}~\bibnamefont{Takubo}},
  \bibinfo{author}{\bibfnamefont{T.}~\bibnamefont{Mizokawa}},
  \bibinfo{author}{\bibfnamefont{M.}~\bibnamefont{Arita}},
  \bibinfo{author}{\bibfnamefont{H.}~\bibnamefont{Namatame}},
  \bibinfo{author}{\bibfnamefont{M.}~\bibnamefont{Taniguchi}},
  \bibinfo{author}{\bibfnamefont{N.}~\bibnamefont{Katayama}},
  \bibinfo{author}{\bibfnamefont{M.}~\bibnamefont{Nohara}}, \bibnamefont{and}
  \bibinfo{author}{\bibfnamefont{H.}~\bibnamefont{Takagi}},
  \bibinfo{journal}{Phys. Rev. Lett.} \textbf{\bibinfo{volume}{103}},
  \bibinfo{pages}{026402} (\bibinfo{year}{2009}).

\bibitem[{\citenamefont{Cercellier et~al.}(2007)\citenamefont{Cercellier,
  Monney, Clerc, Battaglia, Despont, Garnier, Beck, Aebi, Patthey, Berger
  et~al.}}]{Cercellier2007}
\bibinfo{author}{\bibfnamefont{H.}~\bibnamefont{Cercellier}},
  \bibinfo{author}{\bibfnamefont{C.}~\bibnamefont{Monney}},
  \bibinfo{author}{\bibfnamefont{F.}~\bibnamefont{Clerc}},
  \bibinfo{author}{\bibfnamefont{C.}~\bibnamefont{Battaglia}},
  \bibinfo{author}{\bibfnamefont{L.}~\bibnamefont{Despont}},
  \bibinfo{author}{\bibfnamefont{M.~G.} \bibnamefont{Garnier}},
  \bibinfo{author}{\bibfnamefont{H.}~\bibnamefont{Beck}},
  \bibinfo{author}{\bibfnamefont{P.}~\bibnamefont{Aebi}},
  \bibinfo{author}{\bibfnamefont{L.}~\bibnamefont{Patthey}},
  \bibinfo{author}{\bibfnamefont{H.}~\bibnamefont{Berger}},
  \bibnamefont{et~al.}, \bibinfo{journal}{Phys. Rev. Lett.}
  \textbf{\bibinfo{volume}{99}}, \bibinfo{pages}{146403}
  (\bibinfo{year}{2007}).

\bibitem[{\citenamefont{Rontani}(2014)}]{Rontani2014}
\bibinfo{author}{\bibfnamefont{M.}~\bibnamefont{Rontani}},
  \bibinfo{journal}{Phys. Rev. B} \textbf{\bibinfo{volume}{90}},
  \bibinfo{pages}{195415} (\bibinfo{year}{2014}).

\bibitem[{\citenamefont{Varsano et~al.}(2017)\citenamefont{Varsano, Sorella,
  Sangalli, Barborini, Corni, Molinari, and Rontani}}]{Varsano2017}
\bibinfo{author}{\bibfnamefont{D.}~\bibnamefont{Varsano}},
  \bibinfo{author}{\bibfnamefont{S.}~\bibnamefont{Sorella}},
  \bibinfo{author}{\bibfnamefont{D.}~\bibnamefont{Sangalli}},
  \bibinfo{author}{\bibfnamefont{M.}~\bibnamefont{Barborini}},
  \bibinfo{author}{\bibfnamefont{S.}~\bibnamefont{Corni}},
  \bibinfo{author}{\bibfnamefont{E.}~\bibnamefont{Molinari}}, \bibnamefont{and}
  \bibinfo{author}{\bibfnamefont{M.}~\bibnamefont{Rontani}},
  \bibinfo{journal}{Nat. Commun.} \textbf{\bibinfo{volume}{8}},
  \bibinfo{pages}{1461} (\bibinfo{year}{2017}).

\bibitem[{\citenamefont{Drut and L{\"{a}}hde}(2009)}]{Drut2009}
\bibinfo{author}{\bibfnamefont{J.~E.} \bibnamefont{Drut}} \bibnamefont{and}
  \bibinfo{author}{\bibfnamefont{T.~A.} \bibnamefont{L{\"{a}}hde}},
  \bibinfo{journal}{Phys. Rev. Lett.} \textbf{\bibinfo{volume}{102}},
  \bibinfo{pages}{1} (\bibinfo{year}{2009}).

\bibitem[{\citenamefont{Reed et~al.}(2010)\citenamefont{Reed, Uchoa, Joe, Gan,
  Casa, Fradkin, and Abbamonte}}]{Reed2010}
\bibinfo{author}{\bibfnamefont{J.~P.} \bibnamefont{Reed}},
  \bibinfo{author}{\bibfnamefont{B.}~\bibnamefont{Uchoa}},
  \bibinfo{author}{\bibfnamefont{Y.~I.} \bibnamefont{Joe}},
  \bibinfo{author}{\bibfnamefont{Y.}~\bibnamefont{Gan}},
  \bibinfo{author}{\bibfnamefont{D.}~\bibnamefont{Casa}},
  \bibinfo{author}{\bibfnamefont{E.}~\bibnamefont{Fradkin}}, \bibnamefont{and}
  \bibinfo{author}{\bibfnamefont{P.}~\bibnamefont{Abbamonte}},
  \bibinfo{journal}{Science} \textbf{\bibinfo{volume}{330}},
  \bibinfo{pages}{805} (\bibinfo{year}{2010}).

\bibitem[{\citenamefont{Stroucken et~al.}(2011)\citenamefont{Stroucken,
  Gr{\"{o}}nqvist, and Koch}}]{Stroucken2011}
\bibinfo{author}{\bibfnamefont{T.}~\bibnamefont{Stroucken}},
  \bibinfo{author}{\bibfnamefont{J.~H.} \bibnamefont{Gr{\"{o}}nqvist}},
  \bibnamefont{and} \bibinfo{author}{\bibfnamefont{S.~W.} \bibnamefont{Koch}},
  \bibinfo{journal}{Phys. Rev. B} \textbf{\bibinfo{volume}{84}},
  \bibinfo{pages}{205445} (\bibinfo{year}{2011}).

\bibitem[{\citenamefont{{De Juan} and Fertig}(2012)}]{DeJuan2012}
\bibinfo{author}{\bibfnamefont{F.}~\bibnamefont{{De Juan}}} \bibnamefont{and}
  \bibinfo{author}{\bibfnamefont{H.~A.} \bibnamefont{Fertig}},
  \bibinfo{journal}{Phys. Rev. B} \textbf{\bibinfo{volume}{85}},
  \bibinfo{pages}{085441} (\bibinfo{year}{2012}).

\bibitem[{\citenamefont{Stroucken et~al.}(2012)\citenamefont{Stroucken,
  Gr{\"{o}}nqvist, and Koch}}]{Stroucken2012}
\bibinfo{author}{\bibfnamefont{T.}~\bibnamefont{Stroucken}},
  \bibinfo{author}{\bibfnamefont{J.~H.} \bibnamefont{Gr{\"{o}}nqvist}},
  \bibnamefont{and} \bibinfo{author}{\bibfnamefont{S.~W.} \bibnamefont{Koch}},
  \bibinfo{journal}{J. Opt. Soc. Am. B} \textbf{\bibinfo{volume}{29}},
  \bibinfo{pages}{A86} (\bibinfo{year}{2012}).

\bibitem[{\citenamefont{Eberlein et~al.}(2008)\citenamefont{Eberlein, Bangert,
  Nair, Jones, Gass, Bleloch, Novoselov, Geim, and Briddon}}]{Eberlein2008}
\bibinfo{author}{\bibfnamefont{T.}~\bibnamefont{Eberlein}},
  \bibinfo{author}{\bibfnamefont{U.}~\bibnamefont{Bangert}},
  \bibinfo{author}{\bibfnamefont{R.~R.} \bibnamefont{Nair}},
  \bibinfo{author}{\bibfnamefont{R.}~\bibnamefont{Jones}},
  \bibinfo{author}{\bibfnamefont{M.}~\bibnamefont{Gass}},
  \bibinfo{author}{\bibfnamefont{A.~L.} \bibnamefont{Bleloch}},
  \bibinfo{author}{\bibfnamefont{K.~S.} \bibnamefont{Novoselov}},
  \bibinfo{author}{\bibfnamefont{A.}~\bibnamefont{Geim}}, \bibnamefont{and}
  \bibinfo{author}{\bibfnamefont{P.~R.} \bibnamefont{Briddon}},
  \bibinfo{journal}{Phys. Rev. B} \textbf{\bibinfo{volume}{77}},
  \bibinfo{pages}{233406} (\bibinfo{year}{2008}).

\bibitem[{\citenamefont{Gass et~al.}(2008)\citenamefont{Gass, Bangert, Bleloch,
  Wang, Nair, and Geim}}]{Gass2008}
\bibinfo{author}{\bibfnamefont{M.~H.} \bibnamefont{Gass}},
  \bibinfo{author}{\bibfnamefont{U.}~\bibnamefont{Bangert}},
  \bibinfo{author}{\bibfnamefont{A.~L.} \bibnamefont{Bleloch}},
  \bibinfo{author}{\bibfnamefont{P.}~\bibnamefont{Wang}},
  \bibinfo{author}{\bibfnamefont{R.~R.} \bibnamefont{Nair}}, \bibnamefont{and}
  \bibinfo{author}{\bibfnamefont{A.~K.} \bibnamefont{Geim}},
  \bibinfo{journal}{Nature Nanotechnology} \textbf{\bibinfo{volume}{3}},
  \bibinfo{pages}{676} (\bibinfo{year}{2008}).

\bibitem[{\citenamefont{Wang et~al.}(2017)\citenamefont{Wang, Liu, and
  Zhang}}]{Wang2017}
\bibinfo{author}{\bibfnamefont{J.~R.} \bibnamefont{Wang}},
  \bibinfo{author}{\bibfnamefont{G.~Z.} \bibnamefont{Liu}}, \bibnamefont{and}
  \bibinfo{author}{\bibfnamefont{C.~J.} \bibnamefont{Zhang}},
  \bibinfo{journal}{Phys. Rev. B} \textbf{\bibinfo{volume}{95}},
  \bibinfo{pages}{075129} (\bibinfo{year}{2017}).

\bibitem[{\citenamefont{Wang and Liu}(2012)}]{Wang2012a}
\bibinfo{author}{\bibfnamefont{J.~R.} \bibnamefont{Wang}} \bibnamefont{and}
  \bibinfo{author}{\bibfnamefont{G.~Z.} \bibnamefont{Liu}},
  \bibinfo{journal}{New Journal of Physics} \textbf{\bibinfo{volume}{14}},
  \bibinfo{pages}{043036} (\bibinfo{year}{2012}).

\bibitem[{\citenamefont{Popovici et~al.}(2013)\citenamefont{Popovici, Fischer,
  and {Von Smekal}}}]{Popovici2013}
\bibinfo{author}{\bibfnamefont{C.}~\bibnamefont{Popovici}},
  \bibinfo{author}{\bibfnamefont{C.~S.} \bibnamefont{Fischer}},
  \bibnamefont{and} \bibinfo{author}{\bibfnamefont{L.}~\bibnamefont{{Von
  Smekal}}}, \bibinfo{journal}{Phys. Rev. B} \textbf{\bibinfo{volume}{88}},
  \bibinfo{pages}{205429} (\bibinfo{year}{2013}).

\bibitem[{\citenamefont{Ulybyshev}(2013)}]{Ulybyshev2013}
\bibinfo{author}{\bibfnamefont{M.}~\bibnamefont{Ulybyshev}},
  \bibinfo{journal}{Phys. Rev. Lett.} \textbf{\bibinfo{volume}{111}},
  \bibinfo{pages}{056801} (\bibinfo{year}{2013}).

\bibitem[{\citenamefont{Smith and {Von Smekal}}(2014)}]{Smith2014}
\bibinfo{author}{\bibfnamefont{D.}~\bibnamefont{Smith}} \bibnamefont{and}
  \bibinfo{author}{\bibfnamefont{L.}~\bibnamefont{{Von Smekal}}},
  \bibinfo{journal}{Phys. Rev. B} \textbf{\bibinfo{volume}{89}},
  \bibinfo{pages}{195429} (\bibinfo{year}{2014}).

\bibitem[{\citenamefont{Gonz{\'{a}}lez}(2015)}]{Gonzalez2015}
\bibinfo{author}{\bibfnamefont{J.}~\bibnamefont{Gonz{\'{a}}lez}},
  \bibinfo{journal}{Phys. Rev. B} \textbf{\bibinfo{volume}{92}},
  \bibinfo{pages}{125115} (\bibinfo{year}{2015}).

\bibitem[{\citenamefont{Carrington et~al.}(2016)\citenamefont{Carrington,
  Fischer, {Von Smekal}, and Thoma}}]{Carrington2016}
\bibinfo{author}{\bibfnamefont{M.~E.} \bibnamefont{Carrington}},
  \bibinfo{author}{\bibfnamefont{C.~S.} \bibnamefont{Fischer}},
  \bibinfo{author}{\bibfnamefont{L.}~\bibnamefont{{Von Smekal}}},
  \bibnamefont{and} \bibinfo{author}{\bibfnamefont{M.~H.} \bibnamefont{Thoma}},
  \bibinfo{journal}{Phys. Rev. B} \textbf{\bibinfo{volume}{94}},
  \bibinfo{pages}{125102} (\bibinfo{year}{2016}).

\bibitem[{\citenamefont{Molle et~al.}(2017)\citenamefont{Molle, Goldberger,
  Houssa, Xu, Zhang, and Akinwande}}]{Molle2017}
\bibinfo{author}{\bibfnamefont{A.}~\bibnamefont{Molle}},
  \bibinfo{author}{\bibfnamefont{J.}~\bibnamefont{Goldberger}},
  \bibinfo{author}{\bibfnamefont{M.}~\bibnamefont{Houssa}},
  \bibinfo{author}{\bibfnamefont{Y.}~\bibnamefont{Xu}},
  \bibinfo{author}{\bibfnamefont{S.~C.} \bibnamefont{Zhang}}, \bibnamefont{and}
  \bibinfo{author}{\bibfnamefont{D.}~\bibnamefont{Akinwande}},
  \bibinfo{journal}{Nat. Mater.} \textbf{\bibinfo{volume}{16}},
  \bibinfo{pages}{163} (\bibinfo{year}{2017}).

\bibitem[{\citenamefont{Tabert and Nicol}(2014)}]{Tabert2014}
\bibinfo{author}{\bibfnamefont{C.~J.} \bibnamefont{Tabert}} \bibnamefont{and}
  \bibinfo{author}{\bibfnamefont{E.~J.} \bibnamefont{Nicol}},
  \bibinfo{journal}{Phys. Rev. B} \textbf{\bibinfo{volume}{89}},
  \bibinfo{pages}{195410} (\bibinfo{year}{2014}).

\bibitem[{\citenamefont{Ezawa}(2012)}]{Ezawa2012}
\bibinfo{author}{\bibfnamefont{M.}~\bibnamefont{Ezawa}}, \bibinfo{journal}{New
  J. Phys.} \textbf{\bibinfo{volume}{14}}, \bibinfo{pages}{033003}
  (\bibinfo{year}{2012}).

\bibitem[{\citenamefont{Ezawa}(2013)}]{Ezawa2013a}
\bibinfo{author}{\bibfnamefont{M.}~\bibnamefont{Ezawa}},
  \bibinfo{journal}{Phys. Rev. Lett.} \textbf{\bibinfo{volume}{110}},
  \bibinfo{pages}{026603} (\bibinfo{year}{2013}).

\bibitem[{\citenamefont{Iurov et~al.}(2017)\citenamefont{Iurov, Gumbs, Huang,
  and Zhemchuzhna}}]{Iurov2017}
\bibinfo{author}{\bibfnamefont{A.}~\bibnamefont{Iurov}},
  \bibinfo{author}{\bibfnamefont{G.}~\bibnamefont{Gumbs}},
  \bibinfo{author}{\bibfnamefont{D.}~\bibnamefont{Huang}}, \bibnamefont{and}
  \bibinfo{author}{\bibfnamefont{L.}~\bibnamefont{Zhemchuzhna}},
  \bibinfo{journal}{J. Appl. Phys.} \textbf{\bibinfo{volume}{121}},
  \bibinfo{pages}{084306} (\bibinfo{year}{2017}).

\bibitem[{\citenamefont{Pulci et~al.}(2012)\citenamefont{Pulci, Gori, Marsili,
  Garbuio, {Del Sole}, and Bechstedt}}]{Pulci2012}
\bibinfo{author}{\bibfnamefont{O.}~\bibnamefont{Pulci}},
  \bibinfo{author}{\bibfnamefont{P.}~\bibnamefont{Gori}},
  \bibinfo{author}{\bibfnamefont{M.}~\bibnamefont{Marsili}},
  \bibinfo{author}{\bibfnamefont{V.}~\bibnamefont{Garbuio}},
  \bibinfo{author}{\bibfnamefont{R.}~\bibnamefont{{Del Sole}}},
  \bibnamefont{and}
  \bibinfo{author}{\bibfnamefont{F.}~\bibnamefont{Bechstedt}},
  \bibinfo{journal}{EPL} \textbf{\bibinfo{volume}{98}}, \bibinfo{pages}{37004}
  (\bibinfo{year}{2012}).

\bibitem[{\citenamefont{Stanley}(1971)}]{Stanley1971}
\bibinfo{author}{\bibfnamefont{H.~E.} \bibnamefont{Stanley}},
  \emph{\bibinfo{title}{{Introduction to Phase Transitions and Critical
  Phenomena}}} (\bibinfo{publisher}{Oxford University Press Inc.},
  \bibinfo{address}{New York}, \bibinfo{year}{1971}), \bibinfo{edition}{1st}
  ed.

\bibitem[{\citenamefont{Rytova}(1967)}]{Rytova1967}
\bibinfo{author}{\bibfnamefont{N.~S.} \bibnamefont{Rytova}},
  \bibinfo{journal}{Proc. MSU Phys., Astron.} \textbf{\bibinfo{volume}{3}},
  \bibinfo{pages}{30} (\bibinfo{year}{1967}),
  \urlprefix\url{https://www.researchgate.net/publication/320224883_Screened_potential_of_a_point_charge_in_a_thin_film}.

\bibitem[{\citenamefont{Keldysh}(1979)}]{Keldysh1979}
\bibinfo{author}{\bibfnamefont{L.~V.} \bibnamefont{Keldysh}},
  \bibinfo{journal}{Sov. Phys. JETP} \textbf{\bibinfo{volume}{29}},
  \bibinfo{pages}{658} (\bibinfo{year}{1979}).

\bibitem[{\citenamefont{Drummond et~al.}(2012)\citenamefont{Drummond,
  Z{\'{o}}lyomi, and Fal'ko}}]{Drummond2012}
\bibinfo{author}{\bibfnamefont{N.~D.} \bibnamefont{Drummond}},
  \bibinfo{author}{\bibfnamefont{V.}~\bibnamefont{Z{\'{o}}lyomi}},
  \bibnamefont{and} \bibinfo{author}{\bibfnamefont{V.~I.}
  \bibnamefont{Fal'ko}}, \bibinfo{journal}{Phys. Rev. B}
  \textbf{\bibinfo{volume}{85}}, \bibinfo{pages}{075423}
  (\bibinfo{year}{2012}).

\bibitem[{\citenamefont{Fadaie et~al.}(2016)\citenamefont{Fadaie,
  Shahtahmassebi, and Roknabad}}]{Fadaie2016}
\bibinfo{author}{\bibfnamefont{M.}~\bibnamefont{Fadaie}},
  \bibinfo{author}{\bibfnamefont{N.}~\bibnamefont{Shahtahmassebi}},
  \bibnamefont{and} \bibinfo{author}{\bibfnamefont{M.~R.}
  \bibnamefont{Roknabad}}, \bibinfo{journal}{Opt. Quant. Electron.}
  \textbf{\bibinfo{volume}{48}}, \bibinfo{pages}{440} (\bibinfo{year}{2016}).

\bibitem[{\citenamefont{Matthes et~al.}(2013)\citenamefont{Matthes, Gori,
  Pulci, and Bechstedt}}]{Matthes2013}
\bibinfo{author}{\bibfnamefont{L.}~\bibnamefont{Matthes}},
  \bibinfo{author}{\bibfnamefont{P.}~\bibnamefont{Gori}},
  \bibinfo{author}{\bibfnamefont{O.}~\bibnamefont{Pulci}}, \bibnamefont{and}
  \bibinfo{author}{\bibfnamefont{F.}~\bibnamefont{Bechstedt}},
  \bibinfo{journal}{Phys. Rev. B} \textbf{\bibinfo{volume}{87}},
  \bibinfo{pages}{035438} (\bibinfo{year}{2013}).

\bibitem[{\citenamefont{Ni et~al.}(2012)\citenamefont{Ni, Liu, Tang, Zheng,
  Zhou, Qin, Gao, Yu, and Lu}}]{Ni2012}
\bibinfo{author}{\bibfnamefont{Z.}~\bibnamefont{Ni}},
  \bibinfo{author}{\bibfnamefont{Q.}~\bibnamefont{Liu}},
  \bibinfo{author}{\bibfnamefont{K.}~\bibnamefont{Tang}},
  \bibinfo{author}{\bibfnamefont{J.}~\bibnamefont{Zheng}},
  \bibinfo{author}{\bibfnamefont{J.}~\bibnamefont{Zhou}},
  \bibinfo{author}{\bibfnamefont{R.}~\bibnamefont{Qin}},
  \bibinfo{author}{\bibfnamefont{Z.}~\bibnamefont{Gao}},
  \bibinfo{author}{\bibfnamefont{D.}~\bibnamefont{Yu}}, \bibnamefont{and}
  \bibinfo{author}{\bibfnamefont{J.}~\bibnamefont{Lu}}, \bibinfo{journal}{Nano
  Lett.} \textbf{\bibinfo{volume}{12}}, \bibinfo{pages}{113}
  (\bibinfo{year}{2012}).

\bibitem[{\citenamefont{Tao et~al.}(2015)\citenamefont{Tao, Cinquanta, Chiappe,
  Grazianetti, Fanciulli, Dubey, Molle, and Akinwande}}]{Tao2015}
\bibinfo{author}{\bibfnamefont{L.}~\bibnamefont{Tao}},
  \bibinfo{author}{\bibfnamefont{E.}~\bibnamefont{Cinquanta}},
  \bibinfo{author}{\bibfnamefont{D.}~\bibnamefont{Chiappe}},
  \bibinfo{author}{\bibfnamefont{C.}~\bibnamefont{Grazianetti}},
  \bibinfo{author}{\bibfnamefont{M.}~\bibnamefont{Fanciulli}},
  \bibinfo{author}{\bibfnamefont{M.}~\bibnamefont{Dubey}},
  \bibinfo{author}{\bibfnamefont{A.}~\bibnamefont{Molle}}, \bibnamefont{and}
  \bibinfo{author}{\bibfnamefont{D.}~\bibnamefont{Akinwande}},
  \bibinfo{journal}{Nat. Nanotechnol.} \textbf{\bibinfo{volume}{10}},
  \bibinfo{pages}{227} (\bibinfo{year}{2015}).

\bibitem[{\citenamefont{Balendhran et~al.}(2015)\citenamefont{Balendhran,
  Walia, Nili, Sriram, and Bhaskaran}}]{Balendhran2015}
\bibinfo{author}{\bibfnamefont{S.}~\bibnamefont{Balendhran}},
  \bibinfo{author}{\bibfnamefont{S.}~\bibnamefont{Walia}},
  \bibinfo{author}{\bibfnamefont{H.}~\bibnamefont{Nili}},
  \bibinfo{author}{\bibfnamefont{S.}~\bibnamefont{Sriram}}, \bibnamefont{and}
  \bibinfo{author}{\bibfnamefont{M.}~\bibnamefont{Bhaskaran}},
  \bibinfo{journal}{Small} \textbf{\bibinfo{volume}{11}}, \bibinfo{pages}{640}
  (\bibinfo{year}{2015}).

\bibitem[{\citenamefont{{Van Der Donck} et~al.}(2017)\citenamefont{{Van Der
  Donck}, Zarenia, and Peeters}}]{VanDerDonck2017}
\bibinfo{author}{\bibfnamefont{M.}~\bibnamefont{{Van Der Donck}}},
  \bibinfo{author}{\bibfnamefont{M.}~\bibnamefont{Zarenia}}, \bibnamefont{and}
  \bibinfo{author}{\bibfnamefont{F.~M.} \bibnamefont{Peeters}},
  \bibinfo{journal}{Phys. Rev. B} \textbf{\bibinfo{volume}{96}},
  \bibinfo{pages}{035131} (\bibinfo{year}{2017}).

\bibitem[{\citenamefont{Fogler et~al.}(2014)\citenamefont{Fogler, Butov, and
  Novoselov}}]{Fogler2014}
\bibinfo{author}{\bibfnamefont{M.~M.} \bibnamefont{Fogler}},
  \bibinfo{author}{\bibfnamefont{L.~V.} \bibnamefont{Butov}}, \bibnamefont{and}
  \bibinfo{author}{\bibfnamefont{K.~S.} \bibnamefont{Novoselov}},
  \bibinfo{journal}{Nat. Commun.} \textbf{\bibinfo{volume}{5}},
  \bibinfo{pages}{4555} (\bibinfo{year}{2014}).

\bibitem[{\citenamefont{Li et~al.}(2013)\citenamefont{Li, Wang, Zhao, and
  Zhao}}]{Li2013}
\bibinfo{author}{\bibfnamefont{L.}~\bibnamefont{Li}},
  \bibinfo{author}{\bibfnamefont{X.}~\bibnamefont{Wang}},
  \bibinfo{author}{\bibfnamefont{X.}~\bibnamefont{Zhao}}, \bibnamefont{and}
  \bibinfo{author}{\bibfnamefont{M.}~\bibnamefont{Zhao}},
  \bibinfo{journal}{Physics Letters A} \textbf{\bibinfo{volume}{377}},
  \bibinfo{pages}{2628} (\bibinfo{year}{2013}).

\bibitem[{\citenamefont{Kaloni et~al.}(2013)\citenamefont{Kaloni, Tahir, and
  Schwingenschl{\"{o}}gl}}]{Kaloni2013a}
\bibinfo{author}{\bibfnamefont{T.~P.} \bibnamefont{Kaloni}},
  \bibinfo{author}{\bibfnamefont{M.}~\bibnamefont{Tahir}}, \bibnamefont{and}
  \bibinfo{author}{\bibfnamefont{U.}~\bibnamefont{Schwingenschl{\"{o}}gl}},
  \bibinfo{journal}{Sci. Rep.} \textbf{\bibinfo{volume}{3}},
  \bibinfo{pages}{3192} (\bibinfo{year}{2013}).

\bibitem[{\citenamefont{Brunetti et~al.}(2018)\citenamefont{Brunetti, Berman,
  and Kezerashvili}}]{Brunetti2018b}
\bibinfo{author}{\bibfnamefont{M.~N.} \bibnamefont{Brunetti}},
  \bibinfo{author}{\bibfnamefont{O.~L.} \bibnamefont{Berman}},
  \bibnamefont{and} \bibinfo{author}{\bibfnamefont{R.~{\relax{Ya}}.}
  \bibnamefont{Kezerashvili}}, \bibinfo{journal}{Phys. Rev. B}
  \textbf{\bibinfo{volume}{98}}, \bibinfo{pages}{125406}
  (\bibinfo{year}{2018}).

\bibitem[{\citenamefont{Korm{\'{a}}nyos
  et~al.}(2015)\citenamefont{Korm{\'{a}}nyos, Burkard, Gmitra, Fabian,
  Z{\'{o}}lyomi, Drummond, and Fal'ko}}]{Kormanyos2015}
\bibinfo{author}{\bibfnamefont{A.}~\bibnamefont{Korm{\'{a}}nyos}},
  \bibinfo{author}{\bibfnamefont{G.}~\bibnamefont{Burkard}},
  \bibinfo{author}{\bibfnamefont{M.}~\bibnamefont{Gmitra}},
  \bibinfo{author}{\bibfnamefont{J.}~\bibnamefont{Fabian}},
  \bibinfo{author}{\bibfnamefont{V.}~\bibnamefont{Z{\'{o}}lyomi}},
  \bibinfo{author}{\bibfnamefont{N.~D.} \bibnamefont{Drummond}},
  \bibnamefont{and} \bibinfo{author}{\bibfnamefont{V.~I.}
  \bibnamefont{Fal'ko}}, \bibinfo{journal}{2D Mater.}
  \textbf{\bibinfo{volume}{2}}, \bibinfo{pages}{022001} (\bibinfo{year}{2015}).

\bibitem[{\citenamefont{Tran et~al.}(2014)\citenamefont{Tran, Soklaski, Liang,
  and Yang}}]{Tran2014}
\bibinfo{author}{\bibfnamefont{V.}~\bibnamefont{Tran}},
  \bibinfo{author}{\bibfnamefont{R.}~\bibnamefont{Soklaski}},
  \bibinfo{author}{\bibfnamefont{Y.}~\bibnamefont{Liang}}, \bibnamefont{and}
  \bibinfo{author}{\bibfnamefont{L.}~\bibnamefont{Yang}},
  \bibinfo{journal}{Phys. Rev. B} \textbf{\bibinfo{volume}{89}},
  \bibinfo{pages}{235319} (\bibinfo{year}{2014}).

\bibitem[{\citenamefont{Liu et~al.}(2014)\citenamefont{Liu, Neal, Zhu, Luo, Xu,
  Tom{\'{a}}nek, and Ye}}]{Liu2014a}
\bibinfo{author}{\bibfnamefont{H.}~\bibnamefont{Liu}},
  \bibinfo{author}{\bibfnamefont{A.~T.} \bibnamefont{Neal}},
  \bibinfo{author}{\bibfnamefont{Z.}~\bibnamefont{Zhu}},
  \bibinfo{author}{\bibfnamefont{Z.}~\bibnamefont{Luo}},
  \bibinfo{author}{\bibfnamefont{X.}~\bibnamefont{Xu}},
  \bibinfo{author}{\bibfnamefont{D.}~\bibnamefont{Tom{\'{a}}nek}},
  \bibnamefont{and} \bibinfo{author}{\bibfnamefont{P.~D.} \bibnamefont{Ye}},
  \bibinfo{journal}{ACS Nano} \textbf{\bibinfo{volume}{8}},
  \bibinfo{pages}{4033} (\bibinfo{year}{2014}).

\bibitem[{\citenamefont{Trushin et~al.}(2016)\citenamefont{Trushin, Goerbig,
  and Belzig}}]{Trushin2016}
\bibinfo{author}{\bibfnamefont{M.}~\bibnamefont{Trushin}},
  \bibinfo{author}{\bibfnamefont{M.~O.} \bibnamefont{Goerbig}},
  \bibnamefont{and} \bibinfo{author}{\bibfnamefont{W.}~\bibnamefont{Belzig}},
  \bibinfo{journal}{Phys. Rev. B} \textbf{\bibinfo{volume}{94}},
  \bibinfo{pages}{041301} (\bibinfo{year}{2016}).

\end{thebibliography}
\end{document}